\documentclass[conference]{IEEEtran}
\IEEEoverridecommandlockouts
\usepackage{subcaption}
\usepackage{fancyhdr,setspace}
\usepackage{comment}
\usepackage{caption}
\usepackage{todonotes}
\usepackage{bm}
\usepackage[compact]{titlesec} 
\usepackage{cite}
\usepackage{xcolor}
\usepackage{blindtext, graphicx, tabularx, multicol, lipsum, subfig,eqnarray,amsmath,chemformula,multirow,threeparttable}
\usepackage{enumitem}
\usepackage[utf8]{inputenc}
\usepackage{amssymb,amsmath}

\usepackage{float}

\usepackage[boxed]{algorithm2e}
\usepackage[noend]{algpseudocode}
\usepackage[compact]{titlesec}
\algnewcommand{\IfThenElse}[3]{% \IfThenElse{<if>}{<then>}{<else>}
  \State \algorithmicif\ #1\ \algorithmicthen\ #2\ \algorithmicelse\ #3}
\makeatletter
\newcommand{\removelatexerror}{\let\@latex@error\@gobble}
\makeatother
\makeatletter
\renewcommand\footnoterule{%
  \kern-3\p@
  \hrule\@width\columnwidth
  \kern2.6\p@}
  \makeatother
\usepackage{letltxmacro}
\usepackage{nicematrix}

\usepackage{amsthm}

\usepackage{amsmath}

\SetKwInput{KwInput}{Input}                % Set the Input
\SetKwInput{KwOutput}{Output}  

\usepackage{tikz}
\usepackage{capt-of}
\usepackage{amssymb}
\usepackage{booktabs} 
\usepackage{pifont}% http://ctan.org/pkg/pifont

\usepackage{diagbox}
\usepackage{xcolor,etoolbox}
\usepackage{dblfloatfix}

\let\mybibitem\bibitem
\renewcommand{\bibitem}[1]{%
\ifstrequal{#1}{edgeTPU}{\color{black}\mybibitem{#1}}
{\ifstrequal{#1}{xyz}{\color{blue}\mybibitem{#1}}
{\color{black}\mybibitem{#1}}}%
}

\ifCLASSINFOpdf
\else
\fi

\usepackage{soul}
\soulregister\grqq7
\soulregister\cite7
\soulregister\autoref7
\soulregister\ref7
\soulregister\texttt7
\soulregister\textit7
\soulregister\textbf7
\soulregister\chipdesign7
\soulregister\glqq7
\soulregister\grqq7

\begin{document}

% Woohoo!!Paper is accepted!!
\title{Hacking the Fabric: Targeting Partial Reconfiguration for Fault Injection in FPGA Fabrics}

% \author{Anonymous Submission
% \\[-5.0ex]
% }

\author{\IEEEauthorblockN{Jayeeta Chaudhuri$^\dagger{}$, Hassan Nassar$^{\ddagger}{}$, Dennis R.E. Gnad$^{\ddagger}{}$, Jörg Henkel$^{\ddagger}{}$, \\
Mehdi B. Tahoori$^{\ddagger}{}$,  and Krishnendu Chakrabarty$^\dagger{}$}
\IEEEauthorblockA{$^\dagger{}$School of Electrical, Computer, and Energy Engineering, Arizona State University, Tempe, AZ, USA \\
$^\ddagger{}$Institute of Computer Engineering, Karlsruhe Institute of Technology (KIT), Karlsruhe, Germany
\\
}
\\[-5.0ex]
}


%Telephone: (800) 555--1212, Fax: (888) 555--1212}


\maketitle
% \thispagestyle{plain}
% \pagestyle{plain}
\begin{abstract}

%Abstract V10
%%Problem intro 
FPGAs are now ubiquitous in cloud computing infrastructures and reconfigurable system-on-chip, particularly for AI acceleration. Major cloud service providers such as Amazon and Microsoft are increasingly incorporating FPGAs for specialized compute-intensive tasks within their data centers. The availability of FPGAs in cloud data centers has opened up new opportunities for users to improve application performance by implementing customizable hardware accelerators directly on the FPGA fabric. \textcolor{black}{However, the virtualization and sharing of FPGA resources among multiple users open up new security risks and threats. We present a novel fault attack methodology capable of causing persistent fault injections in partial bitstreams during the process of FPGA reconfiguration. This attack leverages power-wasters and is timed to inject faults into bitstreams as they are being loaded onto the FPGA through the reconfiguration manager, without needing to remain active throughout the entire reconfiguration process. Our experiments, conducted on a Pynq FPGA setup, demonstrate the feasibility of this attack 
 on various partial application bitstreams, such as a neural network accelerator unit and a signal processing accelerator unit. } %As a suggested countermeasure, we also present a hash-based authentication technique to detect the injected faults in partial bitstreams before FPGA configuration.
% FPGAs are frequently utilized in cloud computing environments for high performance computing and neural network accelerators. Furthermore, multi-tenancy allows multiple users to upload customized modules on the FPGA, while maintaining logical isolation. However, attackers can take advantage of the multi-tenant environment to launch voltage-based attacks and denial-of-service (DoS). An attacker might stealthily split power-wasting ring oscillators (ROs) across multiple windows within an FPGA configuration bitstream, making it challenging for traditional detection mechanisms to identify these dispersed components as part of a larger malicious circuit. We propose a methodology to detect these malicious bitstreams by transforming individual windows within an FPGA bitstream into a graph-based representation. Leveraging this graph structure, our method employs graph convolutional networks (GCNs) in the training phase to capture malicious patterns from the bitstreams. We use the classification
% accuracy, true-positive rate, and false-positive rate metrics to
% quantify the effectiveness of our method across diverse power-wasting circuits on multiple FPGA boards.

\end{abstract}

% \thispagestyle{fancy}
% \fancyhead{}
% \renewcommand{\headrulewidth}{0pt}
% \fancyhf{}
% \fancyfoot[L]{\large Regular Paper}
% \\ 978-1-7281-4823-6/19/\$31.00 \copyright2019 IEEE}
% \fancyfoot[C]{\large INTERNATIONAL TEST CONFERENCE}
% \fancyfoot[R]{\large \thepage}

%----------------------- This is special for ITC --------------------------
\thispagestyle{plain}
% \fi
\IEEEpeerreviewmaketitle

\section{Introduction}
\vspace{-0.1cm}
Field-Programmable Gate-Arrays (FPGAs) are increasingly being used for customized accelerator platforms. Today, FPGAs are widely incorporated by cloud service providers such as Amazon Web Services (AWS) and Microsoft Azure to accelerate user modules in the cloud\cite{ref_17_amazon} \cite{firestone2018azure}.
Due to the flexibility and versatility of FPGAs across a wide range of high-performance computing applications, several industry and academic projects have suggested that FPGAs should operate in multi-tenancy \cite{mbongue2020architecture}\cite{yamakura2021multi}, similar to other shared resources in the cloud (CPUs, GPUs, memory). Specifically, in a multi-tenant scenario, multiple users can simultaneously reconfigure their applications on different modules of the same FPGA. Currently, FPGAs are capable of partial reconfiguration to enable multi-tenancy as it increases utilization and overall efficiency. Logical isolation is
maintained between modules of different users to ensure the
trust in their operation on multi-tenant reconfigurable fabrics \cite{4223233}. 

% A previously demonstrated fault attack focuses on physically accessing the FPGA fabric to perform a power-based side-channel attack to extract the secret key of an encrypted FPGA bitstream \cite{moradi2011vulnerability}.
However, FPGAs, being dynamically reconfigurable, are prone to security vulnerabilities \cite{krautter2018fpgahammer} \cite{SchellenbergG0T18} \cite{zhao2018fpga} \cite{10.1007/s10836-018-5726-9}. \textcolor{black}{In a multi-tenant operation, users may gain access to the partial reconfigurable regions (PRRs) of the FPGA fabric to execute fault attacks by deploying malicious power-wasting circuits; these attacks can lead to severe voltage drops in the FPGA, resulting in computational faults or denial-of-service (DoS) of the FPGA and other co-tenants using it \cite{krautter2018fpgahammer} \cite{7809042} \cite{8844478}.} In \cite{krautter2018fpgahammer}, ring oscillators (ROs) are utilized to create sudden voltage fluctuations, which are then used to carry out fault attacks on the Advanced Encryption Standard (AES) module. \cite{SchellenbergG0T18} describes remote side-channel attacks in a multi-tenant scenario using voltage-based sensors to retrieve the secret key from an AES module. \cite{8457631} shows that remote side-channel attacks can occur by exploiting the crosstalk between long wires, potentially leaking secret information from AES.  

% \cite{zhao2018fpga} presents an RO-based power monitor to capture power traces of a victim module; the power traces are then used to recover the secret key of a cryptographic application implemented by another user on the same FPGA. 

 % The authors in [9-12] have shown examples of glitch amplification circuits and non-combinational ROs that can cause fault injection in cryptographic modules such as the Advanced Encryption Standard (AES), which can induce abrupt voltage fluctuations and cause the FPGA to crash. 
 % In another work, ROs are implemented to cause fault injections on the AES deployed on a multi-tenant FPGA, which results in the generation of incorrect ciphertexts \cite{FPGAhammer}. 

% The incorporation of the reconfiguration manager in a multi-tenant FPGA environment introduces new security vulnerabilities. An attacker may deploy malicious circuits on the multi-tenant FPGA that induce persistent faults in partial bitstreams deployed to the FPGA via the reconfiguration manager. The fault attack causes the bitstreams to perform erroneously when they are configured on the FPGA, or cause the reconfiguration manager to prevent the bitstream from being loaded on to the FPGA, resulting in DoS.
The fault attacks demonstrated in \cite{krautter2018fpgahammer} \cite{7809042} \cite{8844478} require that the power-wasting circuits be activated for the entire duration while the victim module is configured on the FPGA. These attacks  have temporary fault effects on the victim module and typically result in detectable faults that can be identified by prior detection schemes \cite{10.1145/3451236} \cite{9643485}. 

Despite the demonstrated success of previously proposed countermeasures, there is a need to explore new security solutions;  attackers can exploit the vulnerabilities in new techniques for managing dynamic reconfiguration in FPGAs. For example, to efficiently manage bitstream reconfiguration in the PRR of an FPGA, the authors in \cite{7946114} propose a runtime reconfiguration manager (RM), namely Command-based Reconfiguration Queue (CoRQ). The RM can store multiple partial bitstreams that are uploaded by different users, and subsequently perform partial reconfiguration of these bitstreams on the FPGA \cite{9643485}.

\begin{figure}
\centering
\includegraphics[width=0.44\textwidth]{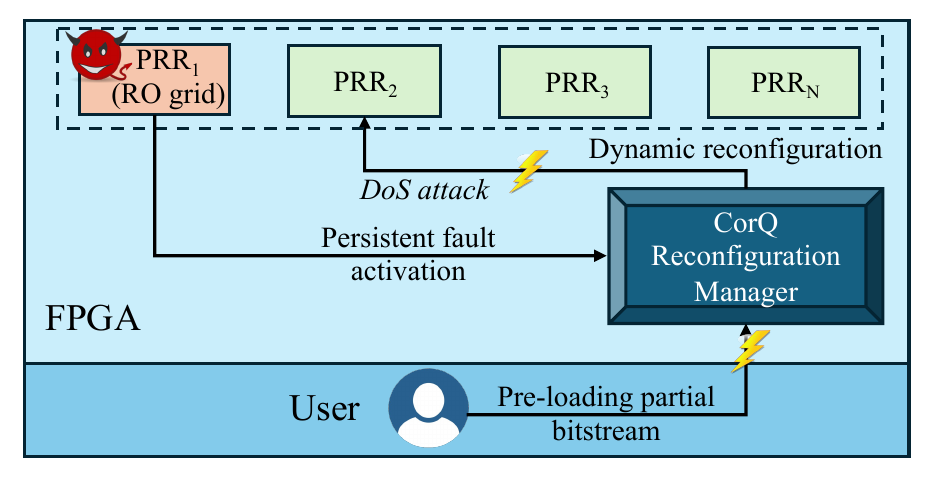}

\caption{Proposed persistent reconfiguration fault attack.}
\label{threat}
\vspace{-0.5cm}
\end{figure}

In this paper, we introduce a
novel type of FPGA-internal fault attack that can be activated
and deactivated remotely. This attack exploits the process through which bitstreams are loaded to the FPGA, specifically targeting the partial reconfiguration process. \textcolor{black}{The attack is precisely timed, e.g., based on sensing side-channel leakages \cite{schellenberg2021inside}, to inject faults in a bitstream \textbf{while} it is being reconfigured on the RM, and cease once the bitstream is fully loaded onto the RM. As highlighted in \cite{10.1145/3451236}, voltage sensors typically take a longer time to detect most attacks. The proposed attack is designed to be active only during the reconfiguration process, which lasts only milliseconds; hence, it evades detection by \cite{10.1145/3451236} \cite{9643485}. }Additionally, the fault introduced during bitstream reconfiguration is not merely transient. Instead, the injected fault becomes embedded within the bitstream itself. Therefore, even after the bitstream is reconfigured to the FPGA, the faults persist in the FPGA. We refer to this type of new attacks as \textit{persistent reconfiguration fault attacks}. Fig. \ref{threat} illustrates the fault-attack scenario.

The key contributions of this paper are as follows.
\begin{itemize}[leftmargin=*,topsep=0pt]
% \item  We develop a calibration algorithm to determine appropriate fault injection parameters for triggering persistent faults in the AES when it is configured within a multi-tenant environment.
\item We show the first persistent reconfiguration fault attack where we leverage pre-configured power-wasters to corrupt user bitstreams while they are being loaded through the RM in a multi-tenant FPGA environment.
\item  We introduce persistent faults into partial bitstreams that configure the neural network accelerator unit and the digital signal processing unit, resulting in erroneous computations after their reconfiguration on the FPGA.
   \item We highlight the effectiveness of our attack through experimental demonstration using a Xilinx FPGA board.
   % \item To detect the proposed attack, we develop a hash-based mechanism within the reconfiguration manager to  verify the authenticity of an user bitstream before reconfiguring it on the FPGA. Using this method, we detect a persistent fault attack and block the fault-injected bitstream from FPGA reconfiguration. 

\end{itemize}

The remainder of the paper is organized as follows. Section II describes prior attacks and related work on multi-tenant FPGAs. Section III presents the proposed attack methodology and the threat model. In Section IV, we demonstrate successful persistent reconfiguration fault attacks on several bitstreams. Section V concludes the paper. 
\vspace{-0.1cm}

% \textcolor{blue}{Arjun}

\section{Background and Related Prior Work}

\vspace{-0.1cm}

\subsection{Security Threats for Multi-Tenant FPGAs}
\vspace{-0.1cm}

In a multi-tenant environment, multiple users share the same chip, but run multiple contexts, which often means that they also share the same power distribution network (PDN). The PDN is responsible for supplying power to all the user modules located on the FPGA. It is composed of a network of resistive, capacitive, and inductive elements. 
While higher resistance leads to a greater  voltage drop, rapid current fluctuations can cause significant voltage spikes due to inductance. In \cite{krautter2018fpgahammer}, authors show fault-injection attacks via power-wasting RO circuits on an AES module deployed on a multi-tenant FPGA. The voltage fluctuations induced in the PDN are impacted by the following parameters of the RO grid: (a) toggling frequency, (b) number of active cycles when the RO grid is enabled, and (c) number of RO instances. Due to voltage fluctuations from the ROs, timing fault injection is made possible, which makes the AES module generate faulty ciphertexts. These ciphertexts are then analyzed using Differential Fault Analysis (DFA) to retrieve the secret key. 

In \cite{8056840}, the authors demonstrate that ROs occupying less than 12\% of the LUTs on the FPGA fabric are sufficient to cause excessive voltage fluctuations of the FPGA. Activating a grid of ROs to switch on and off at a specific frequency, namely $f_{toggle}$ may lead to high power consumption, potentially causing voltage fluctuations. These voltage fluctuations can cause voltage-based faults or crashes. In \cite{gnad2020remote}, voltage-based fault attacks can be observed when the percentage of LUTs configured by ROs is as low as 25\%. 

% Therefore, ROs are a major threat to multi-tenant FPGAs due to their ability to induce fault injection, voltage-based attacks, and DoS. 

\begin{figure}

\includegraphics[width=0.5\textwidth]{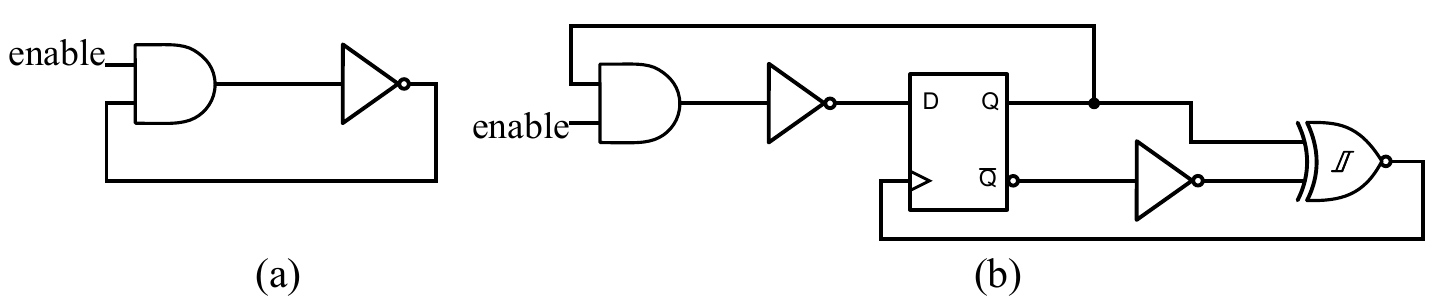}

\caption{(a) Single instance of an RO, (b) Self-clocked RO.}
\label{mal}
\vspace{-0.4cm}
\end{figure}

In \cite{luo2021deepstrike}, authors present a fault-injection attack on deep neural network (DNN) accelerators running on cloud FPGAs. An on-chip voltage sensor based on time-to-digital converters (TDC) is used to monitor the voltage fluctuations \cite{10.1145/2435264.2435283}, while a malicious power-wasting circuit incurs glitching at specific time instances. High power consumption results in voltage glitches, impacting computation in the DNN layers and resulting in timing faults. The glitch-inducing circuits contain self-oscillating loops, which evade design rule check (DRC). Malicious power-wasting circuits based on AES and shift registers are shown in \cite{provelengios2020power}, while \cite{Sugawara2019OscillatorCentre} explore non-combinational ROs based on latches and flip-flops. When activated at specific frequencies, even non-combinational ROs cause significantly high voltage drops leading to voltage and power-based attacks on the FPGA. 
% For example, a latched RO divides the inverter loop by placing a latch in the combinational loop. On the other hand, a self-clocked RO uses a flip-flop and a XOR gate to induce clock glitches. 
Since these circuits do not contain combinational loops, they evade DRC warnings, and can be maliciously deployed on cloud FPGAs to launch power-based attacks.  The work in \cite{10113856} demonstrates successful bitstream generation corresponding to non-combinational ROs on the AWS platform. Examples of non-combinational circuits that are capable of fault-injection attacks are shown in Fig. \ref{mal}. 

In \cite{8844478}, an attack model is demonstrated that does not require specific types of power-wasting circuits such as combinatorial loops, which might leave detectable signatures in the FPGA bitstream. Instead, the attack leverages the capability of causing short circuits in the FPGA by writing opposite logic values concurrently to the same memory address from both ports of a dual-port RAM in the FPGA. The resulting memory collisions can create severe voltage drop and excessive heating of the FPGA, leading to DoS.
\vspace{-0.1cm}

\subsection{Detecting FPGA-Based Attacks}
\vspace{-0.1cm}
Several countermeasures have been explored to prevent fault attacks based on power-wasting circuits, where some methods check the designs before they are getting loaded, and others try to detect the attack during runtime.

% To already prevent the deployment of attacker designs, %malicious power-wasting circuits on multi-tenant FPGAs.

%[offline methods]
To prevent deployment of malicious designs,
%in \cite{8742262}, a 
reverse engineering (RE)-based techniques are used to convert an FPGA bitstream into its technology-mapped netlist. The netlist is then analyzed for combinational loops and other malicious circuits before deploying the corresponding bitstream on the cloud FPGA \cite{Krautter2019MitigatingCloud}. AWS DRC also has basic capabilities for detecting some of the possible combinational loops within user designs that are deployed on their cloud infrastructure \cite{Sugawara2019OscillatorCentre} \cite{AWS_FPGA}. \cite{La2020FPGADefender:FPGAs} and \cite{la2021denial} explore bitstream checking mechanisms that are applicable to large-scale cloud deployments. The checks for malicious bitstream detection include combinational loops, presence of large fanout, short circuits, and glitch detection. 
% \cite{8891998} this one mentions it but is not about bitstream checking
Machine learning (ML)-based detection methods have also been proposed recently to extract malicious signatures directly from bitstreams. In \cite{9810438}, bitstreams are converted into image files and used to train a convolutional neural network (CNN) model to identify RO-based signatures. \cite{10113856} additionally diagnoses the type of malicious circuit implemented by a specific bitstream. These methods can effectively detect non-combinational ROs as well as several power-wasting circuits. 

To detect fault attacks at runtime, in \cite{10.1145/3451236}, the authors develop an on-chip voltage sensor to monitor voltage fluctuations periodically across several regions (or tenants) of an FPGA. In another work, a RO-based power monitor is used to detect the insertion of power-wasting circuits in the FPGA \cite{6339235}. In \cite{10.1145/2435264.2435283}, voltage-based attacks are detected using TDCs, where in \cite{9643485}, that information is used to disable specific interconnect points of attacker modules, reducing the impact of these attacks.

% 10.1145/2435264.2435283 - zick nanoseconds

%9643485, loopbreaker
 % author={Nassar, Hassan and others},

\section{Persistent Attacks on FPGA Tenants During Reconfiguration}
\vspace{-0.1cm}
\subsection{Proposed Persistent Attack Methodology}
\vspace{-0.1cm}
We leverage the multi-tenancy capability of an FPGA to  induce persistent reconfiguration fault attacks on partial bitstreams. Specifically, 
we use power-wasters that already reside in our malicious PRR (cf. Fig.~\ref{threat}) to integrate persistent faults into user designs while they are being dynamically reconfigured on the PRRs of the FPGA through a RM.  Once activated, the power-wasters induce persistent fault  attacks that corrupt the partial bitstreams being loaded to the FPGA. Contrary to prior fault attacks demonstrated in \cite{krautter2018fpgahammer} \cite{7809042} \cite{8844478}, the persistent reconfiguration fault attack is precisely timed to corrupt the partial bitstream only when it is being loaded to the FPGA. Once the faults are injected, they continuously corrupt the bitstream configured to the PRR of the FPGA, and persist throughout the operation of the FPGA until the device is reconfigured. Consequently, even if the power-wasters are deactivated, the FPGA continues to execute based on the corrupted configuration, producing faulty computations. In summary, the proposed attack is particularly  unique from prior fault attacks presented in \cite{krautter2018fpgahammer} \cite{7809042} \cite{8844478} because it:
\begin{itemize}[leftmargin=*,topsep=0pt]
    \item Reduces the duration for which the power-wasters need to remain activated, referred to as  {\em the exposure window}, thus minimizing the risk of detection by voltage sensors \cite{10.1145/3451236}.
\item Extends the attack impact, as the fault-injected bitstream remains corrupted until the next reconfiguration.

\end{itemize}   

% the duration of power-waster activation (which makes the attack potentially detectable), i.e., {\em the exposure window} is reduced, 2) the attack effect is prolonged and become persistent as the faulty loaded bitstream is permanently corrupted until next reconfiguration. both of these factors make this attack unique and different from all existing FAs [refs].

% Therefore, this attack exploits a critical vulnerability during the reconfiguration phase of partial bitstreams on the FPGA. 

\textcolor{black}{Although the persistent fault attack scenario targets a specific RM in our experiments, it is important to note that the proposed attack is generic and can be applied to a range of different reconfiguration platforms used in FPGAs.} The proposed fault attack results in erroneous computations when these designs are deployed on the FPGA. We show the feasibility and severity of the proposed attack on two classes of typical use-cases for FPGA, specifically neural network acceleration and signal processing. 

\begin{table}[t]
\centering
% \begin{threeparttable}
%r2d2 cap attack current sensing
\fontsize{6}{6}\selectfont

\caption{Comparison of proposed attack framework with prior FPGA-based attacks.}

    \label{compare}
    \begin{tabular}{|c| c| c| c| c|c|} 
    \hline
% \multirow{2}{*}{$K$}&
{Attribute} & \cite{krautter2018fpgahammer}&\cite{8056840}&\cite{7809042}& \cite{8844478}&\textbf{Proposed} \\
&&&&&\textbf{method} \\
\hline
Target&AES &FPGA&FPGA  & RAM &Partial\\
module&&&bitstream&&bitstream\\
&&&&& \\
Attack &RO grid&RO grid&Modifying &Memory &RO grid\\
mechanism&&&LUT values&collisions& \\

\multirow{2}{*}{FPGA} &Cyclone V&Virtex-6,&Virtex-5&Spartan-6&Pynq\\
&&Zynq&&&\\
&&&&& \\
Attack &Key &DoS&Key &Timing faults, &Fault-injection,\\
objective&recovery&&recovery&bit-flips& DoS \\
\multirow{2}{*}{Caused DoS?}&\multirow{2}{*}{Yes}&\multirow{2}{*}{Yes}&\multirow{2}{*}{No}&\multirow{2}{*}{Yes}&\multirow{2}{*}{Yes}\\
&&&&& \\
% Mitigation strategy&\multirow{2}{*}{No}&\multirow{2}{*}{No}&\multirow{2}{*}{No}&\multirow{2}{*}{No}&\multirow{2}{*}{Yes}\\
% implemented?&&&&&\\
% &&&&& \\
\multirow{2}{*}{}Self-clocked RO&\multirow{2}{*}{No}&\multirow{2}{*}{No}&\multirow{2}{*}{No}&\multirow{2}{*}{No}&\multirow{2}{*}{Yes}\\
evaluated?&&&&&\\

\hline
\end{tabular}

\vspace{-0.4cm}

% Add32: adder in 32-bit PE; Mult32: multiplier in 32-bit PE; PE16: 16-bit PE; TF: topological features; $A_c$: classification accuracy; $\Delta_S$: savings in test cost.
% \end{threeparttable}
\label{comp}
\end{table}

Table \ref{comp} presents a qualitative comparison of the proposed fault attack with prior attacks on multi-tenant FPGAs.

\vspace{-0.1cm}
\subsection{Threat Model}
\vspace{-0.1cm}

% The threat model for launching fault attacks on the CoRQ framework is
% illustrated in Fig. \ref{threat}. 
We assume that the FPGA is spatially shared by different user modules. In this multi-tenant scenario, an adversary can be a third-party user who gains access to the FPGA to deploy a malicious circuit.   
A RM is configured on the same FPGA. The RM is responsible for (1) pre-loading the partial bitstreams for FPGA reconfiguration, and (2) configuring the bitstreams in the PRRs of the FPGA.     \textcolor{black}{The RM is considered semi-trusted, meaning that another tenant on the same FPGA could exploit a vulnerability of the RM, e.g., side-channel leakage. }In this scenario, the goal of the attacker is to configure a malicious power-waster on one of the PRRs of the FPGA such that it induces faults while other user bitstreams are getting configured on a PRR of the same FPGA through the RM. \textcolor{black}{The power-waster can be a voltage sensor such as RO, which detects side-channel leakages, such as voltage fluctuations, during the loading of a partial bitstream on the RM \cite{schellenberg2021inside} \cite{meyers2023active}. For instance, by monitoring the voltage levels, an attacker can time the activation of the power-wasters to inject faults into the partial bitstream.} Although the PRRs of the FPGA support logical isolation, the PDN of the FPGA is shared among these modules. When the power-wasters are activated at a particular toggling frequency $f_{toggle}$, it causes a sudden voltage drop in the PDN. Due to the abrupt voltage fluctuations, faults can be injected into the bitstream while it is being reconfigured on the FPGA through the RM.  
% These fault persist even after the design is deployed on the PRR of the FPGA, which leads to erroneous computations. 

% In a multi-tenant scenario, several users can upload their customized modules on the partial reconfigurable regions (PRRs) of the FPGA (marked as $PRR_x$ in Fig. \ref{threat}). All the PRRs share a common power distribution network (PDN). Thus, an attacker co-tenant can attempt to induce a huge power consumption on the FPGA and disrupt the performance of victim PRR modules. As shown in \cite{FPGAhammer}, activating a large number of ROs at a particular frequency is sufficient to cause significant power consumption, causing the FPGA to shut down automatically. In this work, we assume that an attacker deploys malicious stealthy, power-wasting circuits for inducing voltage fluctuations in the PDN of the FPGA, causing DoS. Moreover, an attacker gaining illegitimate access to the placed and routed netlist might embed malicious circuits before bitstream generation. We also consider the scenario where an attacker attempts to alter the bitstream during its transmission, before FPGA deployment. 
The RM enables the loading of both encrypted and decrypted bitstreams. However, bitstreams are usually decrypted before they are configured on the FPGA \cite{bit}. Therefore, we evaluate our proposed attack
specifically on decrypted and unencrypted bitstreams that are loaded to the FPGA. 

\vspace{-0.1cm}

\section{Experimental Platform and Results}
\vspace{-0.1cm}

\subsection{Fundamentals of CoRQ framework}
\vspace{-0.1cm}

\begin{figure}
\centering
\includegraphics[width=0.5\linewidth]{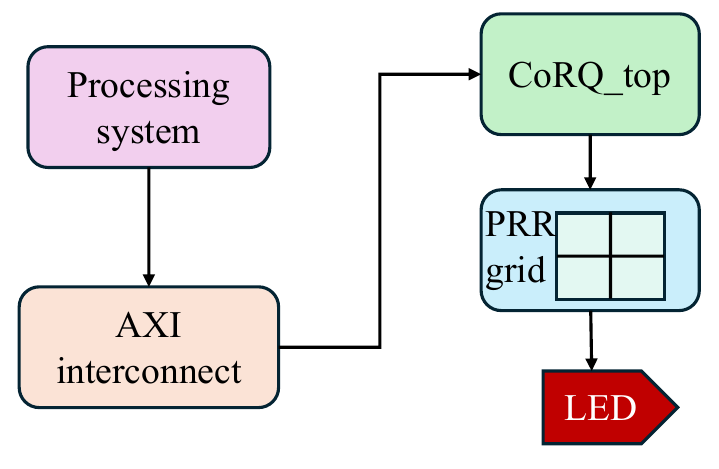}
\caption{Implementation of CoRQ on Pynq FPGA.}
\label{corq}
\vspace{-0.3cm}
\end{figure}

CoRQ \cite{7946114} was implemented as a runtime RM framework for placing partial bitstreams onto the FPGA. If a bitstream is encrypted, CoRQ uses specific commands to decrypt it before FPGA reconfiguration. Most of the FPGA vendors such as Xilinx and Intel decrypt a bitstream during the configuration process \cite{bit}. Therefore, in all our experiments on CoRQ, we evaluate the fault attack using unencrypted partial bitstreams. The RTL implementation of CoRQ is written in VHDL. The block diagram of CoRQ is shown in Fig. \ref{corq}. The modules illustrated in Fig. \ref{corq} are explained below.
\begin{itemize}[leftmargin=*,topsep=0pt]
    \item CoRQ\_top: It denotes the CoRQ reconfiguration framework. CoRQ includes a \textbf{Bitstore} memory and an internal configuration access port (ICAP). All the configuration bitstreams are first stored into the memory, before being reconfigured on the FPGA via the ICAP. 
   
    \item AXI interconnect: The AXI is the standard interface used for facilitating high-bandwidth communication between the processing system and the programmable logic of the FPGA. The use of AXI interfacing in CoRQ facilitates reading the bitstreams from the Bitstore memory and dynamically reconfiguring them on the FPGA. 
    \item Processing system: It denotes the processor core. 
    % \item Blinker blackbox: It controls the LED outputs of the Pynq FPGA according to the design functionality.
\end{itemize}

The application code involving the software driver for using CoRQ and the bitstream decryption units have been implemented in C.
The partial bitstreams (.bin format) are converted into header files using a Python program `\textit{encrypt.py}' before uploading it to the software driver of CoRQ. Embedding the bitstream as a header file enables the bitstream to be included within the application code, thus facilitating dynamic partial reconfiguration. If the evaluated partial bitstreams are unencrypted, the process bypasses any need for decryption.

\begin{figure}
\includegraphics[width=\linewidth]{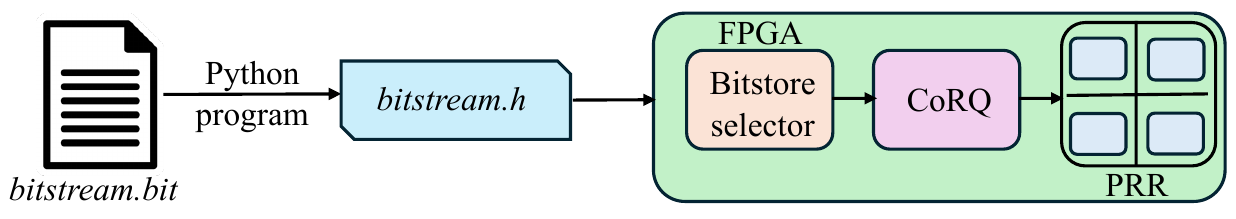}
\caption{Procedure of reconfiguring a partial bitstream on the FPGA through CoRQ.}
\label{corq_process}
\vspace{-0.4cm}
\end{figure}

After implementation of the CoRQ and the grid of ROs on the Pynq FPGA, we export the hardware from Vivado to Vitis. This generates the .XSA (Xilinx Support Archive) file, which helps to run the software application associated with CoRQ. Once uploaded, CoRQ reconfigures the header files corresponding to each partial bitstream on the static PRRs of the FPGA. The procedure of configuring an user bitstream via CoRQ is shown in Fig. \ref{corq_process}. 

% The pseudocode for preloading partial bitstreams via CoRQ is shown in Fig. \ref{alg}.  
% The functions listed in Fig. \ref{alg} are described as follows. 

% \begin{itemize}[leftmargin=*,topsep=0pt]
%     \item \textbf{\textit{test\_bitstream\_decrypt}}: It performs decryption of the encrypted partial bitstreams and reports the number of clock cycles required to decrypt the bitstreams. The decryption time for encrypted partial bitstreams is listed in Table \ref{decrypt}. 
%     \item \textbf{\textit{corq\_timeout}}: This indicates that the decryption of partial bitstreams has been successful, and they are ready to be uploaded to the Bitstore.
%     \item \textbf{\textit{sel\_bitstore}}: This function is implemented as a mux, which allows bitstreams to be written into the Bitstore. 
%     \item \textbf{\textit{corq\_run\_test\_bitstream}}: It enables fetching of a partial bitstream from the Bitstore, preloading it to the CoRQ, and configuring the bitstream on a PRR of the FPGA. 
% \end{itemize}

While a bitstream is being uploaded to the FPGA through CoRQ, we monitor the $corq\_status$ flag to detect a persistent fault-injection attack during the reconfiguration process.
\vspace{-0.1cm}
% \begin{figure}[t]
% \begin{algorithm}[H]
% \fontsize{7}{0}{
% \SetAlgoLined
% \footnotesize\\
%     \KwInput{ \fontsize{8}{8}{}Test bitstream $T$ $\in$ \{\textit{blinkall}, \textit{blinkline}, \textit{blinkcount}, \textit{mac}, \textit{fft}\} }
%     \KwOutput{$corq\_status$  } /*Indicates if bitstream is successfully preloaded on CoRQ*/ % , Error value list $E$}
%     \\
%  \If{$T$ is encrypted} {
%   $flag$ = \textbf{\textit{test\_bitstream\_decrypt}}($T$) \\
%   \If{$flag$ ==`true'}
%   {\textbf{\textit{corq\_timeout}}()} \\
  
%   \Else {print(`Test failed')}}
% \textbf{\textit{sel\_bitstore}}($T$) /*Prepare to preload decrypted/unencrypted bitstreams to CoRQ*/ \\

%   $corq\_status$ = \textbf{\textit{corq\_run\_test\_bitstream}}($T$) /*Preload bitstream on CoRQ*/ \\
%   \If{$corq\_status$ == `false'}
%   {print(`Error in loading bitstream')}
%   \Else{print(`Successfully loaded bitstream')} 
%   /*Ready for FPGA configuration*/

%    \Return $corq\_status$ \\

% \end{algorithm}

% \caption{Pseudocode for preloading partial bitstreams to CoRQ via software driver.}\label{alg}

% \vspace{-0.5cm}
% % \end{algorithm} 

% \end{figure}

\subsection{Fault-Attack Evaluation}
\vspace{-0.1cm}
% In this subsection, we evaluate the impact of a persistent reconfiguration fault attack using power-wasters on partial bitstreams that are uploaded to an FPGA through CoRQ. 
We present three case studies to evaluate the persistent reconfiguration fault-injection attack. In the first case study, we monitor the blinking of the LEDs of the FPGA to detect initiation of fault injection in a partial design. For this experiment, we evaluate the following partial bitstreams -- blinkall, blinkline, and blinkcount. The above bitstreams are implemented specifically to test their effect on the blinking of the FPGA LEDs. The blinkall bitstream enables blinking of all the LEDs simultaneously. The blinkline bitstream causes the LEDs to blink in a sequence. Finally, the blinkcount bitstream involves blinking LEDs in a pattern that represents a count sequence. In the second case study, we study persistent reconfiguration fault attacks on a Multiply-Accumulate unit, which is a crucial component in neural network accelerators that are deployed by cloud providers such as AWS F1 instance. In the third case study, we further demonstrate the fault attack on a Fast Fourier Transform unit. The Fast Fourier Transform unit is crucial for many applications such as speech and audio recognition and analysis, and for polynomial multiplication of homomorphic encryption which run on cloud \cite{jcp3010004}.
Several commercially available FPGAs such as Intel Altera and Xilinx Zynq support Multiply-Accumulate and Fast Fourier Transform operations for performing high-performance computing applications such as deep learning and signal processing. 

% Hence, we perform fault injection on these units while they are deployed in a multi-tenant setup.

In the following case studies, we present fault-attack results where the attacker \textcolor{black}{uses power-wasters based on ROs to disrupt the partial configuration of designs on the multi-tenant FPGA, such that persistent errors are there in the configured design.}  We choose the number of RO instances as 50,000 to be consistent with experimental setups in prior work \cite{gnad2020remote}. To initiate the fault-injection process during the loading of user designs to the RM, we activate the \textit{ro\_ena} signal in the power-waster. Enabling the \textit{`ro\_ena'} bit activates the power-wasters to perform fault-injection during loading of the partial bitstreams to CoRQ. A demo of the fault-attack experiments is available in \cite{demo}, showing errors in the computation of a neural network accelerator unit and a digital signal processing unit when power-wasters are activated. The fault attack is evaluated on a Pynq-Z1 FPGA.

\vspace{-0.1cm}

\begin{table}
\centering
\caption{Decryption time of encrypted partial bitstreams.}
\fontsize{7.2}{7.2}\selectfont

\begin{tabular}{|c |c |c|} 
\cline{1-3}
{Type of encryption circuit}&Size&Time\\ 
  &(in bytes)&(in clock cycles)\\
 
\hline

% {aes\_small}& 80& 2358\\

aes\_medium&65532&533820\\

aes\_big &65540&534006\\
aes\_full &131068&1066358\\
% ascon\_small &80&2096\\
% ascon\_medium &65532&560666 \\
% ascon\_big &65540&560586 \\
% ascon\_full &131068&1119640 \\
\hline
\end{tabular}
 \vspace{-0.2cm}
 \label{decrypt}

%%%%add perf. eval
\end{table}

\subsubsection{Case Study 1: Fault Attack Observed via Blinking LED}
\vspace{-0.1cm}

We evaluate the persistent reconfiguration fault attack on the blinkall, blinkline, and blinkcount partial bitstreams. We convert these bitstreams to their corresponding header files before preloading them to CoRQ.
 Table \ref{blink} lists the size of the header files corresponding to these bitstreams.
\begin{table}
\centering
\caption{The size (in bytes) of the evaluated partial bitstreams.}
%\fontsize{8.2}{8.2}\selectfont
\small
\begin{tabular}{|c |c |c |c |} 
\cline{1-4}
Bitstream & blinkline & blinkall & blinkcount \\

\hline
Size (in bytes) & 51956 & 40524 & 40820 \\

\hline
\end{tabular}
 \vspace{-0.4cm}
 \label{blink}

%%%%add perf. eval
\end{table}

In common practice, a cyclic redundancy check (CRC) is enabled by default in the partial bitstreams (in this specific case, blinkline, blinkall, and blinkcount) before they are configured on the FPGA \cite{crc}. When enabled, an initial CRC value is generated for the original partial bitstream. When the partial bitstream is loaded to CoRQ, the CRC value is recalculated before it is configured on the FPGA. If the recalculated CRC value does not match the original value, it is flagged as a fault-injection attack. Therefore, if a fault occurs during dynamic partial reconfiguration of an user design, CoRQ will raise a flag, indicating that one or multiple bits in the partial bitstream is corrupted. As a result, CoRQ will stop the partial reconfiguration process for that specific design. In cases when the partial bitstream is encrypted, the CRC is disabled. The CRC can be disabled during the bitstream generation process using the constraint \textit{BITSTREAM.GENERAL.CRC Disable} in Xilinx Vivado. For this particular case study, we report our observation by 1) enabling and 2) disabling CRC. 

\textbf{Enabling CRC} We keep CRC enabled during the bitstream generation process of the blinkall, blinkline, and blinkcount designs. After bitstream generation, we obtain the header files corresponding to the bitstreams as explained in Section IV.A. Next, we launch Vitis IDE through Vivado to access the software controller application of CoRQ, preload the partial header files to CoRQ, and then dynamically reconfigure them on the FPGA. While uploading the blinkall bitstream to CoRQ, an error is observed in $corq\_status$ as follows: 

\textit{Error while loading blinkall\_plain. corq\_return: 2} 

Thus, when the CRC is enabled, the persistent reconfiguration fault attack is immediately detected at the preloading stage, which prevents the design from being configured to the FPGA; this attack is demonstrated in \cite{demo}. From \cite{demo}, we observe that only if all the bitstreams are successfully preloaded to CoRQ, they are reconfigured to the FPGA. Therefore, if the fault attack occurs in one of the partial bitstreams, the remaining bitstreams are also prevented from FPGA configuration. This leads to a DoS of CoRQ. 

\textbf{Disabling CRC} Next, we generate the partial bitstreams by disabling the CRC during Vivado bitstream generation. By disabling the CRC, we ensure that no verification occurs when preloading the bitstreams to CoRQ, allowing the fault-injected bitstreams to be loaded onto the FPGA via CoRQ. A demo of the glitching of LEDs induced by the persistent reconfiguration fault attack is shown in \cite{demo}. \textcolor{black}{We perform $200$ iterations of the preloading of the blinkline, blinkcount, and blinkall bitstreams to statistically assess the impact of RO-induced persistent faults on the FPGA LED behavior.} We observe that the blinkline and blinkcount bitstreams are not impacted by the persistent fault-injection attack using ROs. Therefore, when these specific bitstreams are configured to the FPGA via CoRQ, the LEDs blink according to their corresponding functionalities. For example, when the blinkline is uploaded to the FPGA, the LEDs blink in a sequence, indicating that it has been successfully configured. However, the blinkall bitstream is injected by the faults. Therefore, when the faulty blinkall bitstream is uploaded to the FPGA, we observe glitching in the rhythm of blinking of the LEDs on the FPGA.

\vspace{-0.2cm}

% \begin{figure}
% \centering
% \includegraphics[width=0.6\linewidth]{Figures/corq_error.pdf}
% \caption{Terminal output while reconfiguring partial bitstreams via CoRQ (a) Indicates decryption has been performed successfully, (b) Decrypted/unencrypted partial bitstreams are loaded into the Bitstore, (c) In the absence of malicious ROs, CoRQ successfully loads the partial bitstreams to the FPGA (d) CoRQ is unable to load the partial bitstream `blinkall\_plain', indicating that a fault attack has occurred.}
% \label{attack_corq}
% \end{figure}

\subsubsection{Case Study 2: Fault Attack on a Neural Network Accelerator Unit}
\vspace{-0.1cm}
Next, we demonstrate the persistent reconfiguration fault attack on a neural network accelerator component, specifically the Multiply-Accumulate unit. The partial bitstream that implements the Multiply-Accumulate unit is converted to its corresponding header file, known as \textit{mac.h}. This file, \textit{mac.h} is then loaded onto the FPGA using CoRQ for evaluation. We provide a pair of test inputs ($v_1, v_2$) to the Multiply-Accumulate unit configured on the FPGA, and collect the output from the Multiply-Accumulate operation. We specify the register addresses, namely \textit{IN\_A} and \textit{IN\_B} to store the test input values $v_1$ and $v_2$, respectively. \textcolor{black}{After the Multiply-Accumulate computation, we read the data from the output port of the Pynq FPGA. }The computed value received from the FPGA, is given by $N_{comp}$. The expected value $N_{mac}$ is given by: $N_{mac} = N_{mac}^{'} + v_1\times v_2$, where $N_{mac}^{'}$ represents the previously accumulated value.

We use the metric $e_{mac}$ to quantify the error between the computed output and the expected outcome. The normalized error $e_{mac}$ (absolute value) is given by $e_{mac}=|\frac{N_{mac}-N_{comp}}{N_{mac}}|$, where $N_{mac}$ is the expected output of the Multiply-Accumulate operation and $N_{comp}$ is the computed value in the presence of a persistent reconfiguration fault attack. We choose the test inputs randomly from the range [1, 10], and perform the Multiply-Accumulate computation over 10 iterations i.e., for 10 different test input pairs ($v_1, v_2$). The partial accumulated sum from $(i-1)^{th}$ iteration is added to the product of test input pair ($v_1, v_2$) in the $i^{th}$ iteration.

\begin{figure}
\centering
\includegraphics[width=0.85\linewidth]{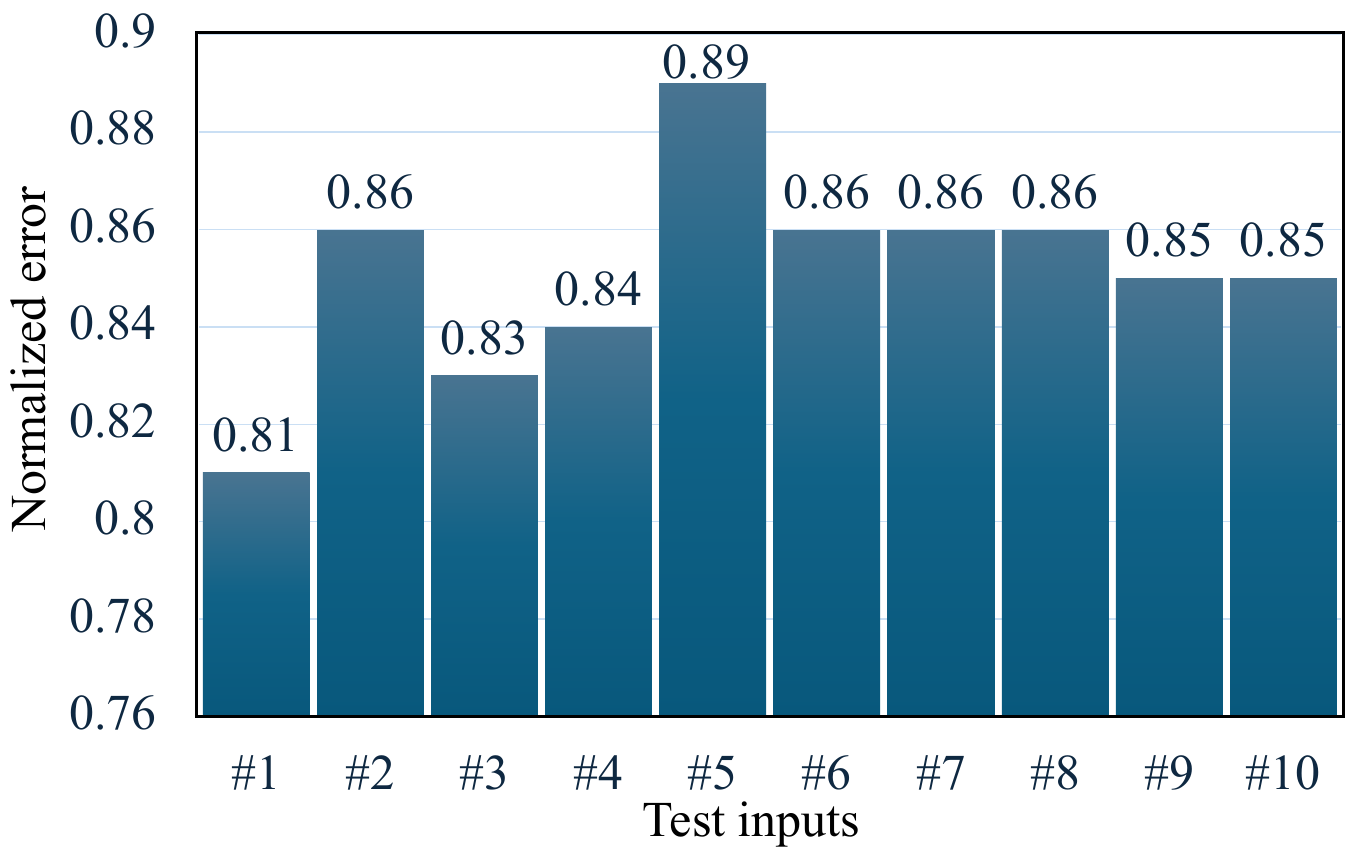}

\caption{Impact of fault injection (in terms of the normalized error) on a partial user design implementing the Multiply-Accumulate unit.}
\label{mac}
\vspace{-0.5cm}
\end{figure}

Fig. \ref{mac} illustrates the normalized error due to fault-injection attack on the Multiply-Accumulate unit. Table \ref{compute} lists the $N_{comp}$ and $e_{mac}$ values over 1000 iterations; the faulty $N_{comp}$ outputs  highlight the severity of RO-based fault-injection.
\vspace{-0.1cm}

\subsubsection{Case Study 3: Fault Attack on a Digital Signal Processing Unit}
\vspace{-0.1cm}

\begin{figure}
\centering
\includegraphics[width=0.43\textwidth]{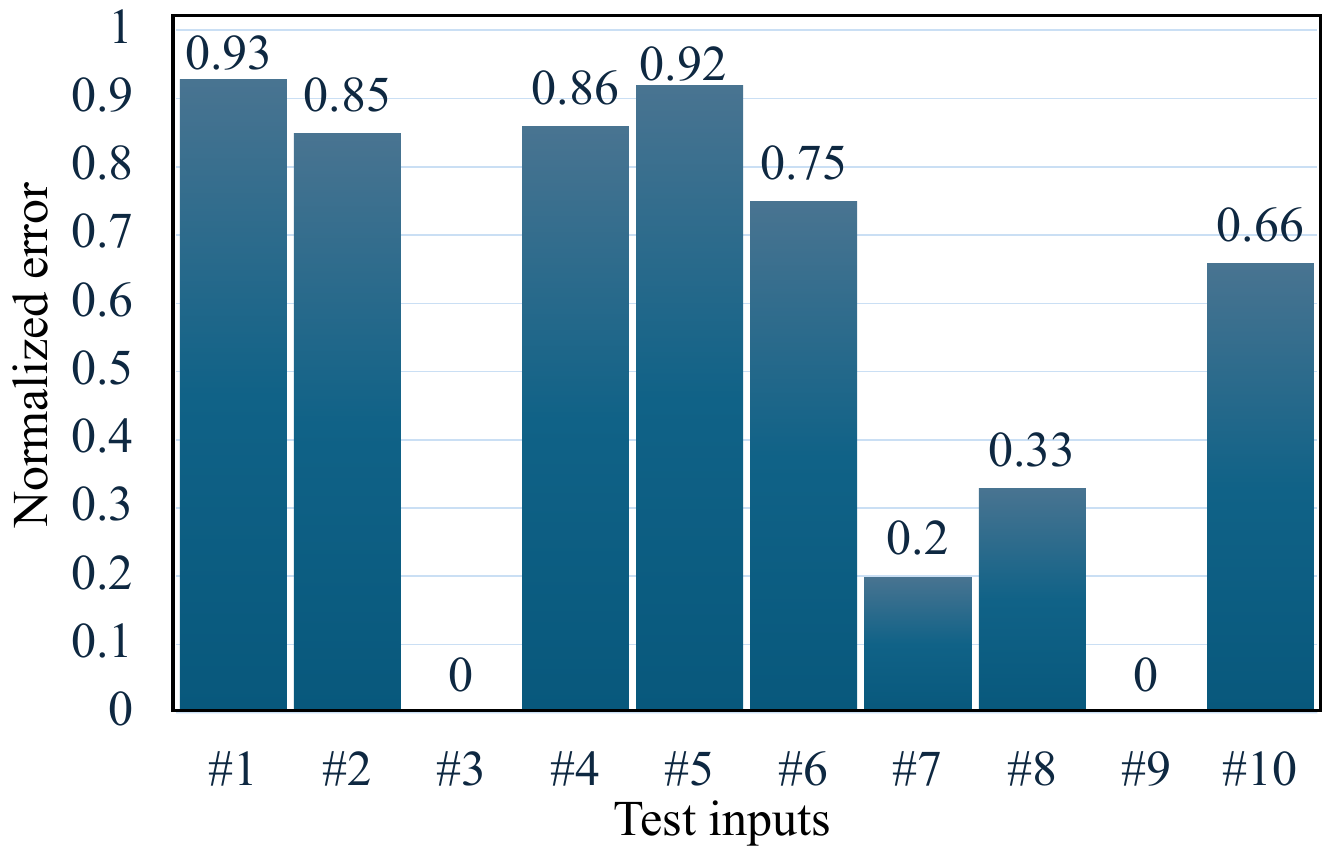}

\caption{Impact of fault injection (in terms of the normalized error) on a partial bitstream implementing the Fast Fourier Transform unit.}
\label{fft}
\vspace{-0.5cm}
\end{figure}

We show evaluation results of the persistent reconfiguration fault attack for a digital signal processing component, specifically the Fast Fourier Transform unit. The partial bitstream implementing Fast Fourier Transform is converted to the respective header file, \textit{fft.h}. As explained in the previous subsection, we use a similar technique to provide a pair of test inputs ($v_1, v_2$) to the Fast Fourier Transform unit that is configured on the FPGA, and read the computed data from the FPGA. The actual (expected) outcome of the Fast Fourier Transform operation, $N_{fft}$ is given by $N_{fft} = v_1\times v_2$. The normalized error $e_{fft}$ (absolute value) is formulated as $e_{fft}=|\frac{N_{fft}-N_{comp}}{N_{fft}}|$, where $N_{fft}$ is the expected output of the Fast Fourier Transform computation and $N_{comp}$ is the computed value in the presence of the persistent reconfiguration fault attack. For the $i^{th}$ iteration, test inputs are chosen randomly from the range [0, 10]; the computation in each iteration is independent of the previous iterations. Fig. \ref{fft} shows the normalized error over 10 different test input pairs. We  report the $e_{fft}$ values over 1000 iterations in Table \ref{compute}.

\vspace{-0.1cm}

\begin{table}
\centering
\caption{Normalized error for fault-injected Multiply-Accumulate and Fast Fourier Transform units.}
\fontsize{8}{8}\selectfont

\begin{tabular}{|c |c |c|c|c|c|c|} 
\cline{1-5}
{Iteration}&\multicolumn{2}{c|}{Multiply-Accumulate}& \multicolumn{2}{c|}{Fast Fourier Transform}\\
 
\hline
&$N_{comp}$&$e_{mac}$&$N_{comp}$&$e_{fft}$ \\
\cline{2-5}
10&39&0.84&5&0.66\\

50 &343&0.76&9&0.87\\

100 &875&0.66&13&0.62 \\
250 &1967&0.73&10&0.66\\
500  &3648&0.76&0&1\\
750  &5584&0.75&6&0.7\\
1000  &7673&0.74&12&0.88\\

\hline
\end{tabular}
 % \vspace{-0.2cm}
 \label{compute}

%%%%add perf. eval
\end{table}

\subsection{Comparison to Runtime Fault-Injection Attacks}
\vspace{-0.1cm}

Several previous works have discussed fault injection at runtime.
\cite{krautter2018fpgahammer} used ISCAS-based circuits to inject faults on AES.
\cite{gnad2020remote} explored the right combinations of fault-injection parameters for causing fault attacks on specific types of FPGAs.
\cite{boutros20} targets injecting faults to DNNs.
All these methods require the attack to persist for the whole runtime of the victim accelerator to be effective. 
For example, if the victim accelerator runs for two hours, the attack will have to run for two hours.
This duration puts the attack at risk as simple monitoring of the power activity can give it away.
In contrast, our attack induces persistent faults and only needs to be active during the upload of the bitstream, typically in the range of milliseconds.
Once, the bitstream is configured to the PRR of the FPGA, the attack can cease, yet the faults remain embedded in the configured bitstream.
Therefore, even if our attack is detected, it will already have affected the accelerator and the fault will persist.

\vspace{-0.1cm}

\section{Conclusion}
\vspace{-0.1cm}
We have demonstrated persistent fault attacks on several partial bitstreams that are uploaded to the FPGA via the RM, and have shown how these attacks can successfully lead to faulty computations in a multi-tenant environment. We have shown that for as low as 15\% of FPGA LUTs used for implementing the power-wasters, the RM is unable to load partial bitstreams on the FPGA, resulting in DoS. Additionally, we have demonstrated successful fault attacks on several partial bitstreams while they are being loaded to the RM, thus highlighting a critical vulnerability of the FPGA reconfiguration process.

\vspace{-0.1cm}

% \section*{Acknowledgement}

% The work of Jayeeta Chaudhuri and Krishnendu Chakrabarty was supported in part by the National Science Foundation under grant no. CNS-2011561. The work of Hassan Nassar was supported in part by the German Federal Ministry of Education and Research (BMBF) through grant 01IS23066 as part of the Software Campus Project ``HE-Trust'' and in part by the ``Helmholtz Pilot Program for Core Informatics (kikit)'' at KIT.
% % \vspace{-0.1cm}

% \textcolor{blue}{Arjun}

\bibliographystyle{IEEEtran}

{
\hyphenpenalty=10000
\exhyphenpenalty=10000
\sloppy

\bibliography{references}

\begin{thebibliography}{10}
\providecommand{\url}[1]{#1}
\csname url@samestyle\endcsname
\providecommand{\newblock}{\relax}
\providecommand{\bibinfo}[2]{#2}
\providecommand{\BIBentrySTDinterwordspacing}{\spaceskip=0pt\relax}
\providecommand{\BIBentryALTinterwordstretchfactor}{4}
\providecommand{\BIBentryALTinterwordspacing}{\spaceskip=\fontdimen2\font plus
\BIBentryALTinterwordstretchfactor\fontdimen3\font minus \fontdimen4\font\relax}
\providecommand{\BIBforeignlanguage}[2]{{%
\expandafter\ifx\csname l@#1\endcsname\relax
\typeout{** WARNING: IEEEtran.bst: No hyphenation pattern has been}%
\typeout{** loaded for the language `#1'. Using the pattern for}%
\typeout{** the default language instead.}%
\else
\language=\csname l@#1\endcsname
\fi
#2}}
\providecommand{\BIBdecl}{\relax}
\BIBdecl

\bibitem{ref_17_amazon}
Amazon, ``Amazon {EC}2 {F}1 {I}nstance,'' \url{https://go.aws/3ENtUj9}, 2021.

\bibitem{firestone2018azure}
D.~Firestone \emph{et~al.}, ``Azure accelerated networking: {S}mart{NIC}s in the public cloud,'' in \emph{15th USENIX (NSDI)}, 2018, pp. 51--66.

\bibitem{mbongue2020architecture}
J.~M. Mbongue \emph{et~al.}, ``Architecture support for {FPGA} multi-tenancy in the cloud,'' in \emph{ASAP}.\hskip 1em plus 0.5em minus 0.4em\relax IEEE, 2020, pp. 125--132.

\bibitem{yamakura2021multi}
M.~Yamakura \emph{et~al.}, ``A multi-tenant resource management system for multi-{FPGA} systems,'' \emph{IEICE TRANSACTIONS on Information and Systems}, vol. 104, no.~12, pp. 2078--2088, 2021.

\bibitem{4223233}
T.~Huffmire \emph{et~al.}, ``Moats and drawbridges: An isolation primitive for reconfigurable hardware based systems,'' in \emph{SP '07}, 2007, pp. 281--295.

\bibitem{krautter2018fpgahammer}
J.~Krautter \emph{et~al.}, ``{FPGA}hammer: {R}emote voltage fault attacks on shared {FPGA}s, suitable for {DFA} on {AES},'' \emph{IACR TCHES}, 2018.

\bibitem{SchellenbergG0T18}
F.~Schellenberg \emph{et~al.}, ``An inside job: Remote power analysis attacks on {FPGA}s,'' in \emph{DATE}, 2018.

\bibitem{zhao2018fpga}
M.~Zhao \emph{et~al.}, ``{FPGA}-based remote power side-channel attacks,'' \emph{Proc. IEEE S\&P}, 2018.

\bibitem{10.1007/s10836-018-5726-9}
R.~Elnaggar and K.~Chakrabarty, ``Machine learning for hardware security: Opportunities and risks,'' \emph{JETTA}, vol.~34, no.~2, 2018.

\bibitem{7809042}
P.~Swierczynski \emph{et~al.}, ``Bitstream fault injections ({B}i{FI})–automated fault attacks against {SRAM}-based {FPGA}s,'' \emph{IEEE TC}, 2018.

\bibitem{8844478}
M.~M. Alam \emph{et~al.}, ``{RAM}-{J}am: Remote temperature and voltage fault attack on {FPGA}s using memory collisions,'' in \emph{FDTC}, 2019, pp. 48--55.

\bibitem{8457631}
C.~Ramesh \emph{et~al.}, ``{FPGA} side channel attacks without physical access,'' in \emph{FCCM}, 2018, pp. 45--52.

\bibitem{10.1145/3451236}
G.~Provelengios \emph{et~al.}, ``Mitigating voltage attacks in multi-tenant {FPGA}s,'' \emph{ACM TRETS}, vol.~14, no.~2, 2021.

\bibitem{9643485}
H.~Nassar \emph{et~al.}, ``Loop{B}reaker: Disabling interconnects to mitigate voltage-based attacks in multi-tenant {FPGA}s,'' in \emph{ICCAD}, 2021.

\bibitem{7946114}
M.~Damschen \emph{et~al.}, ``Co{RQ}: Enabling runtime reconfiguration under {WCET} guarantees for real-time systems,'' \emph{IEEE Embedded Systems Letters}, vol.~9, no.~3, pp. 77--80, 2017.

\bibitem{schellenberg2021inside}
F.~Schellenberg \emph{et~al.}, ``An inside job: Remote power analysis attacks on {FPGA}s,'' \emph{IEEE Design \& Test}, vol.~38, no.~3, pp. 58--66, 2021.

\bibitem{8056840}
D.~Gnad \emph{et~al.}, ``Voltage drop-based fault attacks on {FPGA}s using valid bitstreams,'' in \emph{FPL}, 2017, pp. 1--7.

\bibitem{gnad2020remote}
D.~R. Gnad \emph{et~al.}, ``Remote electrical-level security threats to multi-tenant {FPGA}s,'' \emph{IEEE Design \& Test}, vol.~37, no.~2, 2020.

\bibitem{luo2021deepstrike}
Y.~Luo \emph{et~al.}, ``Deepstrike: Remotely-guided fault injection attacks on {DNN} accelerator in cloud-{FPGA},'' in \emph{DAC}, 2021, pp. 295--300.

\bibitem{10.1145/2435264.2435283}
K.~M. Zick \emph{et~al.}, ``Sensing nanosecond-scale voltage attacks and natural transients in {FPGA}s,'' ser. FPGA.\hskip 1em plus 0.5em minus 0.4em\relax ACM, 2013.

\bibitem{provelengios2020power}
G.~Provelengios \emph{et~al.}, ``Power wasting circuits for cloud {FPGA} attacks,'' in \emph{FPL}, 2020.

\bibitem{Sugawara2019OscillatorCentre}
T.~Sugawara \emph{et~al.}, ``{Oscillator without a combinatorial loop and its threat to FPGA in data centre},'' \emph{Electronics Letters}, 2019.

\bibitem{10113856}
J.~Chaudhuri and K.~Chakrabarty, ``Diagnosis of malicious bitstreams in cloud computing {FPGA}s,'' \emph{IEEE T{CAD}}, vol.~42, no.~11, 2023.

\bibitem{Krautter2019MitigatingCloud}
J.~Krautter \emph{et~al.}, ``{Mitigating Electrical-level Attacks towards Secure Multi-Tenant {FPGA}s in the Cloud},'' \emph{ACM TRETS}, vol.~12, no.~3, 2019.

\bibitem{AWS_FPGA}
{Amazon AWS}, ``{A}mazon {AWS} {FPGA} usage,'' \url{https://bit.ly/2XcGnXy}.

\bibitem{La2020FPGADefender:FPGAs}
T.~M. La \emph{et~al.}, ``{FPGADefender: Malicious self-oscillator scanning for Xilinx UltraScale + {FPGA}s},'' \emph{ACM TRETS}, 2020.

\bibitem{la2021denial}
T.~La \emph{et~al.}, ``Denial-of-service on {FPGA}-based cloud infrastructures—attack and defense,'' \emph{IACR TCHES}, 2021.

\bibitem{9810438}
J.~Chaudhuri and K.~Chakrabarty, ``Detection of malicious {FPGA} bitstreams using {CNN}-based learning,'' in \emph{ETS}, 2022.

\bibitem{6339235}
A.~L. Masle and W.~Luk, ``Detecting power attacks on reconfigurable hardware,'' in \emph{FPL}, 2012, pp. 14--19.

\bibitem{meyers2023active}
V.~Meyers \emph{et~al.}, ``Active and passive physical attacks on neural network accelerators,'' \emph{IEEE Design \& Test}, vol.~40, no.~5, pp. 70--85, 2023.

\bibitem{bit}
Xilinx, ``Ultrascale architecture configuration,'' \url{https://bit.ly/3yyxvQ9 }.

\bibitem{jcp3010004}
R.~Kiesel \emph{et~al.}, ``Potential of homomorphic encryption for cloud computing use cases in manufacturing,'' \emph{Journal of Cybersecurity and Privacy}, vol.~3, no.~1, pp. 44--60, 2023.

\bibitem{demo}
``Demo files for fault attacks on {FPGA},'' \url{https://rb.gy/ce4v1x} [Downloading is necessary for the videos to be played].

\bibitem{crc}
AMD, ``Error testing using {POST\_CRC} in 7 {S}eries {FPGA}s,'' \url{https://rb.gy/3mbsag}.

\bibitem{boutros20}
A.~Boutros \emph{et~al.}, ``Neighbors from hell: Voltage attacks against deep learning accelerators on multi-tenant {FPGAs},'' in \emph{ICFPT}, 2020.

\end{thebibliography}
\vspace{-0.4cm}
}

\end{document}